\documentclass[aps,10pt,prc,tightenlines,twocolumn,twoside,showpacs]{revtex4}
\usepackage{graphicx,amsmath,amssymb}

\newcommand{\eg}{\textit{e.g.}}  

\newcommand{\ie}{\textit{i.e.}}  
\newcommand{\dd}{\mathrm{d}}

\begin{document}

\title{Transverse Momentum Spectra of $J/\psi$ in Heavy-Ion Collisions}

\author{Xingbo Zhao} \author{Ralf Rapp} \affiliation{Cyclotron Institute and
  Physics Department, Texas A\&M University, College Station, TX
  77843-3366, USA}

\date{\today}

\begin{abstract}

  We investigate $J/\psi$ transverse-momentum ($p_t$) distributions and
  their centrality dependence in heavy-ion collisions at SPS and RHIC    
  within the framework of a two-component model,
  which includes (i) primordial production coupled with various
  phases of dissociation, (ii) statistical coalescence of $c$ and
  $\bar{c}$ quarks at the hadronization transition. The suppression of 
  the direct component (i) is calculated by solving a transport 
  equation for $J/\psi$, $\chi_c$ and $\psi^\prime$ in an expanding 
  fireball using momentum dependent dissociation rates in the 
  Quark-Gluon Plasma (QGP). The coalescence component is inferred from 
  a kinetic rate equation with a momentum dependence following from a 
  blast wave approach.
  At SPS energies, where the direct component dominates, the interplay 
  of Cronin effect and QGP suppression results in fair agreement with 
  NA50 $p_t$ spectra. At RHIC energies, the $p_t$ spectra 
  in central $Au$+$Au$ collisions are characterized by a transition from
  regeneration at low $p_t$ to direct production above. At lower
  centralities, the latter dominates at all $p_t$.  

\end{abstract}

\maketitle

\section{Introduction}
\label{sec_intro}
It has long been suggested that the suppression~\cite{Matsui:1986dk}
of $J/\psi$ mesons can be utilized as a probe of Quark-Gluon Plasma 
(QGP) formation in ultrarelativistic heavy-ion collisions (URHICs).
This effect has indeed been observed in $Pb$-$Pb$ collisions at the 
Super Proton Synchrotron (SPS)~\cite{Ramello:2003ig}, consistent with 
theoretical models invoking charmonium dissociation reactions in a 
QGP~\cite{Grandchamp:2001pf,Zhu:2004nw,Karsch:2005nk,Linnyk:2006ti}.    
At collider energies, however, a copious production of $c\bar c$ 
pairs has led to the suggestion that charmonia can be regenerated 
by a coalescence of $c$ and $\bar c$ quarks close to the hadronization
transition~\cite{Grandchamp:2001pf,BraunMunzinger:2000px,Gorenstein:2000ck}. 
The presence of this mechanism is supported by data from the Relativistic
Heavy-Ion Collider (RHIC)~\cite{Adare:2006ns}, where, despite the higher 
temperatures of the putative QGP, the observed suppression is similar to 
SPS energies
(as, \eg, predicted in Ref.~\cite{Grandchamp:2001pf}). 
However, a quantitative assessment of the regeneration and suppression
mechanisms at RHIC has not been achieved yet.
   
Transverse momentum ($p_t$) spectra of charmonia are hoped 
to provide additional means of discrimination. 
In Ref.~\cite{Zhu:2004nw}, the average $\langle p_t^2\rangle$ of 
primordial $J/\psi$'s with QGP suppression has been computed, while in 
Refs.~\cite{Bugaev:2002xx,Andronic:2006ky} the regeneration component 
has been studied within a blast wave description based on thermalized
$J/\psi$ mesons at the QCD phase boundary. 
In Refs.~\cite{Greco:2003vf,Thews:2005vj} 
the impact of non-thermalized $c$ quark distributions on the $p_t$
spectra resulting from recombination has been studied, but no
direct component was accounted for.  In Ref.~\cite{Yan:2006ve} the average
$\langle p_t^2\rangle$ and elliptic flow at RHIC has been calculated 
including both direct and regenerated components using gluo-dissociation
rates with vacuum binding energies for the $J/\psi$ and $\chi_c$.

In the present Letter we provide a comprehensive description of $p_t$
spectra at both SPS and RHIC, including their centrality dependence and 
absolute yields. A proper description of the inclusive
yields vs. centrality turns out to be
particularly important for interpreting $p_t$ spectra in terms of
direct and regeneration contributions. We adopt a previously 
constructed two-component approach~\cite{Grandchamp:2001pf} 
employing inelastic charmonium reaction rates (extended to finite 
3-momentum) which account for reduced 
binding energies in the QGP.  The latter point is essential to allow for 
a realistic  treatment of $\chi_c$ and $\psi'$ states, which are expected
to be measured in the future but also make up 30-40\% of the inclusive
initial $J/\psi$ yield. In the following, we recall the basic 
ingredients of the 2-component model and its extension to finite momentum 
(Section~\ref{sec_2comp}), apply our approach to heavy-ion collisions 
at SPS and RHIC (Section~\ref{sec_spec}) and conclude (Section~\ref{sec_concl}).   

\section{Two-Component Approach}
\label{sec_2comp}
In analogy to the case of light hadrons, one may decompose $p_t$ spectra
of charmonia ($\Psi=J/\psi, \chi_c, \psi'$) in URHICs according to their
production mechanism into hard (high-$p_t$) and soft (low-$p_t$) 
components, 
\begin{equation}
\left.\frac{\dd N_{\Psi}}{p_t\dd
  p_t}\right\vert_{tot}=\left.\frac{\dd N_{\Psi}}{p_t\dd
  p_t}\right\vert_{dir}+
\left.\frac{\dd N_{\Psi}}{p_t\dd p_t}\right\vert_{coal}
 \ ,
\label{dndpt}
\end{equation}
where the direct component (first term) is associated with hard production in
primordial $N$-$N$ collisions, subject to suppression in the subsequent
medium evolution. The soft component (second term) is conceptually simpler than in 
the light sector, based on the notion that $c$ and $\bar c$ quarks are 
exclusively produced primordially, leaving their coalescence as the only 
source of secondary charmonium formation. Since regeneration is governed 
by the phase space density of $c$ and $\bar c$ quarks in the medium, 
it is sensitive to the $c$-quark momentum spectra. 
Indirect measurements of $c$-quark spectra at 
RHIC (via semileptonic decay electrons)~\cite{Adare:2006nq,Abelev:2006db} 
indicate strong rescattering effects which in theoretical 
models~\cite{vanHees:2005wb} imply an approximate thermalization up to 
$c$-quark momenta of $p_t^c\sim2-2.5$~GeV~\cite{Greco:2007sz}. Thus, 
following our earlier developed 2-component 
model~\cite{Grandchamp:2001pf,Grandchamp:2002wp}, we approximate the 
coalescence component with a thermal blast wave description, while direct 
production is computed in a microscopic suppression calculation in QGP and 
hadronic phase, as will be detailed in the remainder of this section. 
Both terms in Eq.~(\ref{dndpt}) are evaluated in the same 
expanding fireball model. 


Let us first address the direct component. Since the charmonium masses 
are much larger than the typical temperature of the medium, a Boltzmann 
transport equation is appropriate to describe the time evolution of 
the phase space distribution, $f_\Psi(\vec{x},\vec{p},\tau)$, 
through the QGP, mixed and hadron gas (HG) 
phase, 
\begin{equation}
p^{\mu}\partial_{\mu}f_\Psi(\vec{x},\vec{p},\tau)
=-E_\Psi \ \Gamma_\Psi(\vec{x},\vec{p},\tau) \ 
f_\Psi(\vec{x},\vec{p},\tau) \ ,
\label{boltz}
\end{equation}
where $E_\Psi=\sqrt{m^2_\Psi+p^2}$ is the energy of $\Psi$ with 
3-momentum modulus $p$, and $\vec x$ is its position in the fireball.  
For simplicity, we constrain our calculation to the longitudinal 
rest frame of the charmonium~\cite{Zhu:2004nw}, \ie, solve the Boltzmann 
equation in (2+1)-dimensions. For the initial charmonium distribution 
we assume a factorization into spatial and momentum dependencies,
$f(\vec{x}_t,\vec{p}_t,\tau_0)=f(\vec{x}_t,\tau_0)\cdot f(\vec{p}_t,\tau_0)$, 
where $\tau_0$ is the initial (thermalization) time of the medium (QGP or mixed
phase). The spatial part of the initial distribution is obtained from 
a Glauber model including nuclear absorption,  
\begin{eqnarray}
f_\Psi(\vec{x}_t,\tau_0)=\sigma^\Psi_{pp}\int d^2s\ dz\ dz^{\prime}\rho_A(\vec{s},z)\
\rho_B(\vec{x}_t-\vec{s},z^{\prime})\nonumber\\ \times\exp\left\{-\int^\infty_z 
dz_A\rho_A(\vec{s},z_A)\sigma_{nuc}\right\}\nonumber\\
\times\exp\left\{-\int^\infty_{z^\prime} 
dz_B\rho_B(\vec{x}_t-\vec{s},z_B)\sigma_{nuc}\right\}\ ,
\label{fx_glauber}
\end{eqnarray}
where $\rho_{A,B}$ are Woods-Saxon profiles~\cite{De Jager:1974dg}
of nuclei $A$ and $B$ and $\sigma^\Psi_{pp}$ is the $\Psi$
production cross section in $p$+$p$ collisions (we use 
$d\sigma^{J/\psi}_{pp}/dy(y=0)$=25(750)~nb at SPS 
(RHIC)~\cite{Abreu:1997jh,Adare:2006kf}). The nuclear absorption 
cross section, $\sigma_{nuc}$, serves as a parameter to account
for pre-equilibrium charmonium suppression due to
primordial nucleons passing by, estimated from $p$-$A$
collisions; at SPS we adopt the values of Refs.~\cite{Sitta:2004hj,Borges:2005ab}, 
$\sigma_{nuc}$=4.4~mb for $J/\psi$, $\chi_c$ and 7.9~mb for $\psi'$.  
At RHIC, we use $\sigma_{nuc}$=1.5~mb (for $J/\psi$, $\chi_c$) 
based on Ref.~\cite{Adler:2005ph}, 
which is compatible with a recent update~\cite{Adare:2007gn} (for 
$\sigma_{nuc}$=2.7~mb the total $J/\psi$ yield in our model decreases
by 8$\%$ for central $Au$-$Au$ at RHIC). For $\psi'$, 
we employ an accordingly increased value of 2.7~mb. 
The initial momentum spectra are obtained from $p$+$p$ data, augmented
by a Gaussian smearing to simulate nuclear $p_t$-broadening (Cronin effect), 
\begin{equation}
\label{fp_i}
f_\Psi(\vec{p}_t,\tau_0)=\frac{1}{2\pi\sigma^2}\int\ d^2p^\prime_t\ \exp
(-\frac{p^{\prime2}_t}{2\sigma^2})\ f_{NN}(|\vec{p_t}-\vec{p_t}^\prime|)
\ ,
\end{equation}
where $f_{NN}(p_t)$ is the spectrum in elementary $N$-$N$ collisions. 
At SPS, $f_{NN}(p_t)=\frac{1}{\pi\ \langle
  p^2_t\rangle}\exp(-p^2_t/\langle p^2_t\rangle)$ with $\langle
p^2_t\rangle$=1.15 $\mathrm{GeV}^2/c^2$~\cite{Abreu:2000xe,Topilskaya:2003iy}, and
at RHIC $f_{NN}(p_t)=A\ (1+p^2_t/B^2)^{-6}$ with $B$=4.1~GeV$^2$ 
yielding $\langle p^2_t\rangle_{pp}$=4.14~GeV$^2$~\cite{Adare:2006kf}. 
The Cronin effect is computed using $2\sigma^2$=$a_{gN}\cdot\langle l\rangle$ 
where $\langle l\rangle$ represents the centrality dependent mean nuclear 
path length of the gluons before fusing into $\Psi$~\cite{Hufner:2001tg}.
At SPS, the extracted coefficient is 
$a_{gN}$=0.076GeV$^2$/fm~\cite{Abreu:2000xe,Topilskaya:2003iy}, while a 
fit to $d$-$Au$ data at RHIC gives $a_{gN}$$\simeq$0.1~GeV$^2$/fm
with a rather large uncertainty (\eg, $a_{gN}$=0.6~GeV$^2$/fm 
is still compatible with $d$+$Au$ data, but results in
$R_{AA}$($p_t$=6~GeV)$\approx$9 before QGP suppression in 
0-20$\%$ central $Au$+$Au$).

\begin{figure}[!t]
\hspace*{-4mm}
\includegraphics[width=0.4\textwidth]{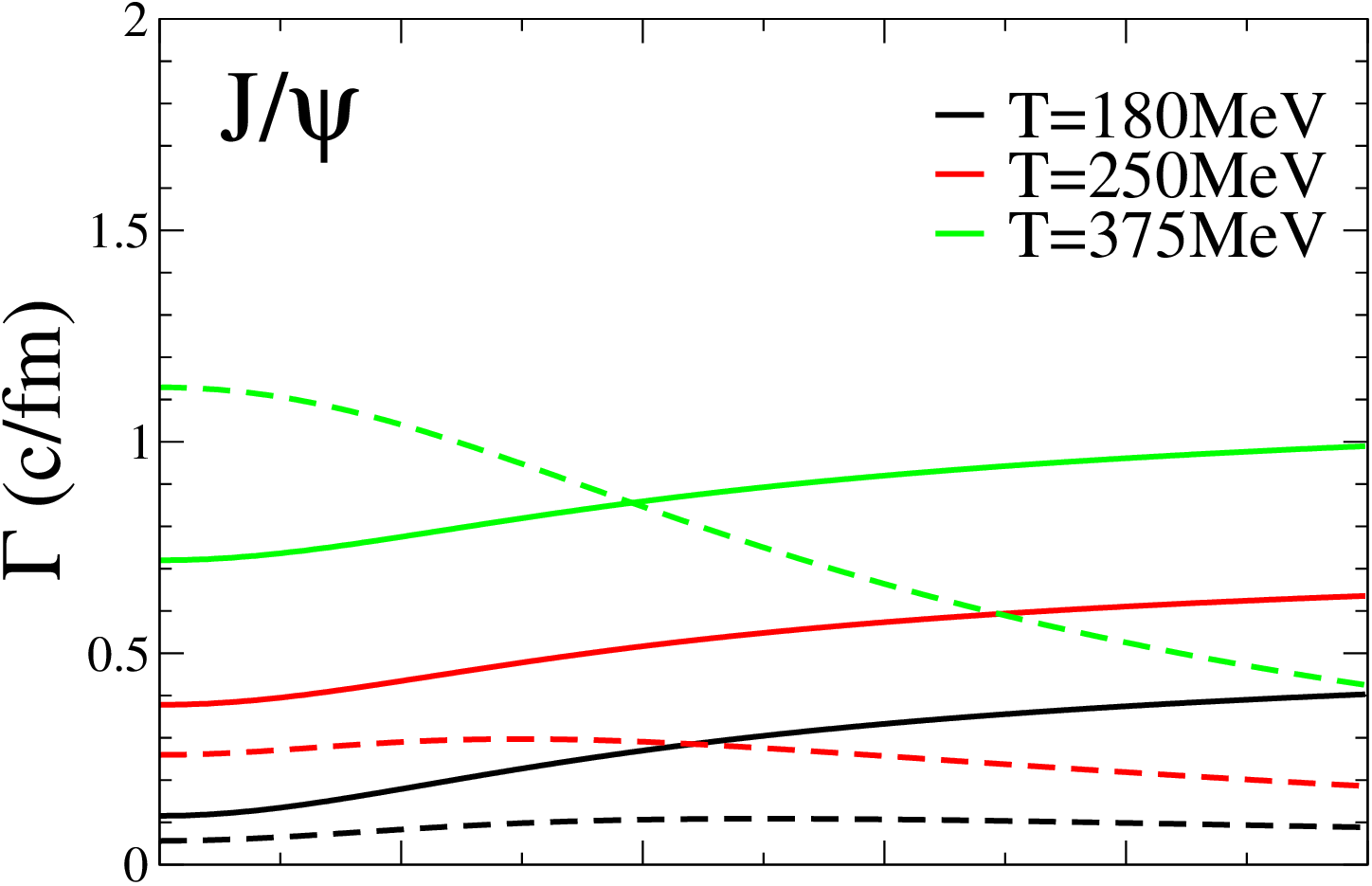}
\includegraphics[width=0.4\textwidth]{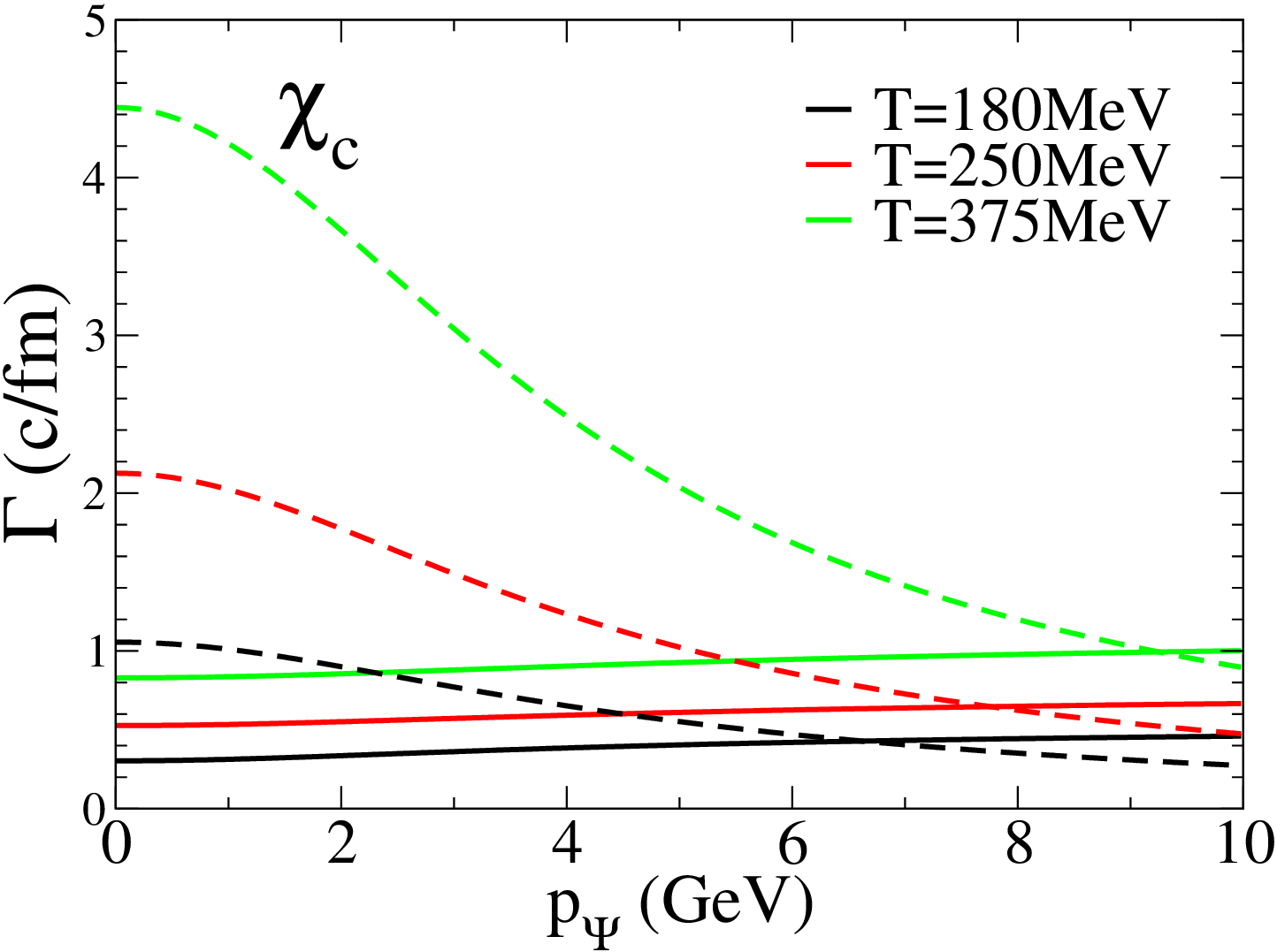}
\caption{(Color online) Comparison of the momentum dependence of
  quasifree (solid lines) and gluo-dissociation rates (dashed lines) 
  for $J/\psi$ (upper panel) and $\chi_c$ (lower panel) at different
  temperatures.} 
\label{fig_rate}
\end{figure}

The most important microscopic ingredient to the transport Eq.~(\ref{boltz}) 
are the charmonium dissociation rates, $\Gamma_\Psi$,
which can be expressed via inelastic cross sections, $\sigma^{diss}_{\Psi i}$, 
for $\Psi$ scattering on medium constituents $i$ as
\begin{equation}
\label{gamma}
\Gamma_\Psi(\vec{x},\ \vec{p},\ \tau) 
= \sum_{i} \int \frac{d^3k}{(2\pi)^3} \ f^i(\omega_k;T(\tau)) 
 \  \sigma^{diss}_{\Psi i} \ v_{rel}\ 
\end{equation}
with $v_{rel}=F/(E_\Psi E_{i})$,  $E_i=(k^2+m^2_i)^{1/2}$, 
flux factor $F=((p^\mu k_\mu)^2-m^2_\Psi m^2_k)^{1/2}$, 
$k_\mu$: parton/meson 4-momentum, $f^i(\omega_k;T)$: thermal Fermi/Bose 
distribution.
In the QGP, color Debye screening is expected to reduce charmonium binding 
energies, $\epsilon_B$ (which eventually vanish), which is supported by recent lattice
QCD calculations~\cite{Aarts:2007pk,Kaczmarek:2007pb}. Under these 
circumstances, gluo-dissociation reactions, $\Psi+g\to c+\bar c$, become 
inefficient and should be replaced by quasifree dissociation, 
$i + \Psi \to i+c+\bar c$ ($i$=$g$,$q$,$\bar q$)~\cite{Grandchamp:2001pf}.
Here we extend these calculations to finite 3-momentum and compare 
the rates to gluo-dissociation (using vacuum binding energies and 
vanishing thermal gluon mass)~\cite{Peskin:1979va} in 
Fig.~\ref{fig_rate}.
The strong coupling constant $\alpha_s$ in the quasifree cross section 
is one of two adjustable parameters in our approach and is fixed to
reproduce the $J/\psi$ yield in central $Pb$-$Pb$ collisions at the SPS
(resulting in $\alpha_s$=0.24).
Except for $J/\psi$'s at rather low temperatures and 3-momenta, 
the gluo-dissociation rate decreases with increasing $p$ 
due to a pronounced maximum in the pertinent cross section (at 
a gluon energy $\omega\simeq1.43\epsilon_B$ in the rest system of the 
$J/\psi$~\cite{Rapp:2005rr}). On the other hand, the quasifree rate 
always increases with $p$  
due to a smoothly increasing cross section with center-of-mass energy,
similar to Ref.~\cite{Song:2007gm}.
This reiterates the importance of using the quasifree rate (rather than
gluo-dissociation) for small binding energies, especially at finite 
momentum. The increase of the quasifree rate with $p$ is more pronounced 
at low temperature since at high temperature most partons are 
energetic enough to destroy a $J/\psi$ irrespective of its momentum.  
This trend is weaker for the $\chi_c$ (lower panel) due to its small 
binding energy: even at low temperature most 
partons carry sufficient energy to destroy it.

In the HG, we employ inelastic cross sections with $\pi$ and $\rho$ mesons 
from a flavor-$SU(4)$ effective Lagrangian 
approach~\cite{Lin:1999ad,Haglin:2000ar}. 

The final ingredient required to solve the transport equation~(\ref{boltz})
is the space-time and temperature evolution of the system, which we 
model by an isentropically expanding fireball model represented by a 
cylindrical volume, 
\begin{equation}
\label{Vfb}
V_{FB}(\tau)=(z_0+v_z\tau+\frac{1}{2} a_z\tau^2) \ \pi \ 
(r_0+\frac{1}{2}a_\perp \tau^2)^2\ ,
\end{equation}
where $z_0$, $v_z$, $a_z$, $r_0$ and $a_\perp$ are the
initial longitudinal length, longitudinal expansion velocity and
acceleration, initial transverse radius and transverse acceleration,
respectively. At fixed total entropy (matched to the observed hadron
multiplicity at given centrality and collision energy), the temperature of
the system follows from the equation of state (massive partons in QGP
and resonance gas in HG). We update the transverse acceleration to
$a_\perp$=0.08~$c^2$/fm and 0.1~$c^2$/fm at SPS and RHIC as used in 
recent applications of the fireball model to 
dilepton~\cite{vanHees:2006ng} and heavy-quark~\cite{vanHees:2005wb}
observables. The initial temperature for central $A$-$A$
collisions at SPS (RHIC) is 210(370)~MeV, with thermal freezeout
at $T_{\rm fo}\simeq110$~MeV.

The Boltzmann transport equation~(\ref{boltz}) can now be solved for 
the final phase-space distribution $f_\Psi(\vec{x}_t,\vec{p}_t,\tau_f)$ 
of $J/\psi$, $\chi_c$ and  $\psi^\prime$ 
at the freeze-out time $\tau_f$ for fixed centrality and collision 
energy.  Upon integration over the transverse plane of the medium we 
obtain  the $p_t$ spectrum of the direct component as
\begin{equation}
\label{fp_d}
\left.\frac{\dd N_{\Psi}}{p_t\dd p_t}\right\vert_{dir} =\int\
d^2x_t\  f(\vec{x}_t,\ \vec{p}_t,\ \tau_f) \ ,
\end{equation}
It is worth noting that the
leakage effect~\cite{Zhuang:2003fu} is implemented 
by switching off the suppression if a charmonium state moves outside 
the fireball, \ie, $\Gamma_\Psi(\vec{x},\ \vec{p},\
\tau)\equiv$0 for $x_t(\tau)>r_0+\frac{1}{2}a_\perp\tau^2$. As we will
see below this effect is significant for charmonia at high $p_t$.

Let us now turn to the coalescence component, \ie, the second term on the 
right-hand side of Eq.~(\ref{dndpt}). As stated above, we assume the 
regenerated charmonia to follow a local thermal equilibrium distribution
with transverse flow velocity given by the blastwave 
expression~\cite{Schnedermann:1993ws},
\begin{equation}
\label{bw}
\left.\frac{\dd N_{\Psi}}{p_t\dd p_t}\right\vert_{coal}\propto 
m_t\int^R_0 rdr K_1\left(\frac{m_t\cosh y_t}{T}\right)
I_0\left(\frac{p_t\sinh y_t}{T}\right) \ 
\end{equation}
($m_t$=$\sqrt{m_\Psi^2+p^2_t}$).
Since charmonium regeneration is inoperative in the 
HG~\cite{Grandchamp:2003uw,Rapp:2005rr}, we evaluate the blast wave formula
at the hadronization transition as following from the fireball model, 
Eq.~(\ref{Vfb}), with $T$=$T_c$=170(180)~MeV and transverse 
rapidity $y_t$=$\tanh^{-1}v_t(r)$ using a
linear flow profile $v_t(r)$=$v_{s}\frac{r}{R}$ with surface velocity 
$v_{s}$=0.33(0.49)$c$ and transverse radius $R$=7.3(7.9)~fm
for central collisions at SPS (RHIC).
To determine the normalization of the coalescence component we utilize 
a momentum-independent rate equation~\cite{Grandchamp:2003uw},
\begin{equation}
\label{rate-eq}
\frac{\dd N_{\Psi}}{\dd
  \tau}=-\Gamma_{\Psi} \ (N_{\Psi}-N_{\Psi}^{\text{eq}})
 \ ,
\end{equation}
where $\Gamma_\Psi\stackrel{\wedge}{=}\Gamma_\Psi(p=0)$ and $N_{\Psi}^{\text{eq}}$ is 
the equilibrium number of charmonia for a given number of $c\bar c$ pairs
in the system (based on total cross sections 
$\sigma_{pp}^{c\bar c}$=5.5(570)~$\mu$b at SPS (RHIC)). The charmonium yield 
due to the gain term is identified with the abundance of the coalescence 
component. As in Refs.~\cite{Grandchamp:2002wp,Grandchamp:2003uw} a 
thermal relaxation time, $\tau^{\rm therm}_c$, for charm quarks
is introduced to mimic a reduced charmonium equilibrium
limit due to incomplete kinetic equilibration via  
a relaxation factor ${\cal R}$=1-$\exp(-\int\ d\tau /\tau^{\rm therm}_c)$. 
Due to the current uncertainties in $\sigma_{pp}^{c\bar c}$, 
$\tau^{\rm therm}_c$ and 
its schematic implementation we cannot quantitatively predict
the coalescence yields. Therefore we adjust $\tau^{\rm therm}_c$ to the 
inclusive $J/\psi$ yield in central $Au$-$Au$ at RHIC. This point will be 
improved in future work by solving the rate equation at finite $p$ with 
(time-dependent) $c$-quark momentum distributions as obtained from 
Langevin simulations~\cite{vanHees:2005wb} which result in fair 
agreement with the semileptonic single-electron $R_{AA}$ and $v_2$ 
at RHIC~\cite{Adare:2006nq,Abelev:2006db}. 
Here, we employ $\tau^{\rm therm}_c$=7~fm/$c$ (in line with the microscopic
approach of Ref.~\cite{van Hees:2007me}), compared to $\sim$2-4~fm/$c$ in
Ref.~\cite{Grandchamp:2003uw}. This update reduces the regeneration
yield by 30-50$\%$ in central collisions at SPS and RHIC (see 
Sec.~\ref{sec_spec} below).

\section{$J/\psi$ Yields and Spectra at SPS and RHIC}
\label{sec_spec}
\begin{figure}[!t]
\includegraphics[width=0.4\textwidth,height=0.32\textwidth]{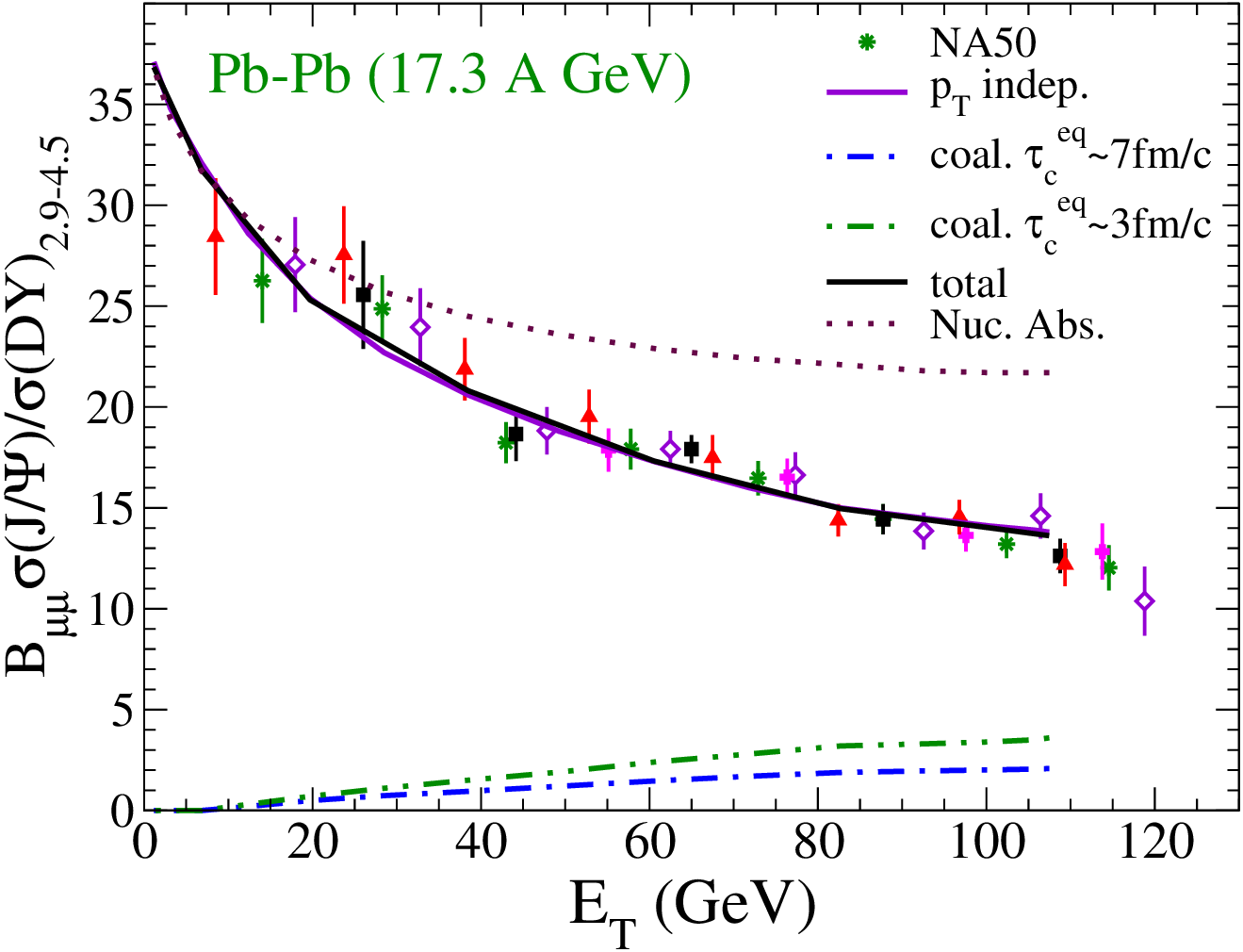}
\caption{(Color online). NA50 data~\cite{Ramello:2003ig} for the 
  centrality dependence of $J/\psi$/Drell-Yan dimuons at SPS  
  compared to our results with (black line) and without (purple line)
  3-momentum dependence (both lines essentially coincide). 
   The sensitivity of the coalescence component 
  to the charm-quark relaxation time
  is indicated by the dash-dotted ($\tau^{\rm therm}_c$=7~fm/$c$) and
  dash-double-dotted ($\tau^{\rm therm}_c$=3~fm/$c$) lines.}
\label{fig_RAASPS}
\end{figure}
We start our phenomenological analysis at SPS energies. The centrality
dependence of inclusive $J/\psi$ production (including feeddown) is 
summarized in Fig.~\ref{fig_RAASPS}. To recover previous results 
from the momentum-independent calculations for central collisions,
a minor reduction of $\alpha_s$ from 0.26 to 0.24 in the quasifree 
dissociation rate has been applied (since the rate increases with $p$).  
After this adjustment, there is no visible modification in the
inclusive centrality dependence left compared to the previous $p$=0
results~\cite{Grandchamp:2003uw} (an increase of the leakage of $J/\psi$'s
for smaller system sizes is essentially compensated by a decrease
in the fireball lifetime).
The sensitivity to the coalescence contribution is small, but this
will be different at RHIC. 
\begin{figure}[!t]
\includegraphics[width=0.4\textwidth]{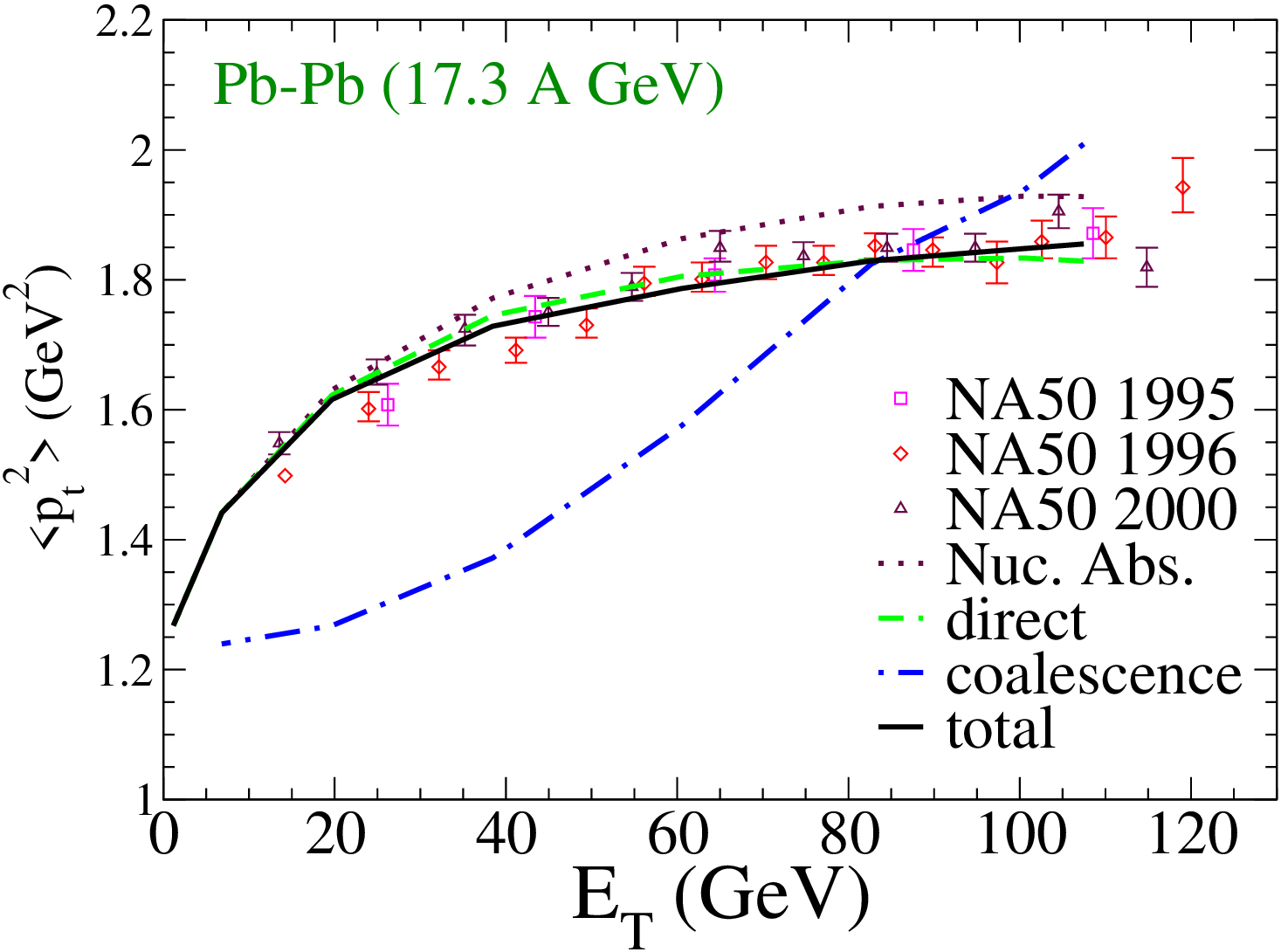}
\caption{(Color online). $\langle p^2_t\rangle$ as a function of
  centrality at SPS. NA50 data~\cite{Abreu:2000xe,Topilskaya:2003iy} are 
  compared to our model calculations. The $\langle p^2_t\rangle$ for the
  direct component (dashed line) and coalescence component (dash-dotted
  line) are compared to the $\langle p^2_t\rangle$ with nuclear absorption 
  only including Cronin $p_t$-broadening effect (dotted line).}
\label{fig_pt2SPS}
\end{figure}
Our calculated $J/\psi$ $p_t$ spectra are used to compute its average
$\langle p_t^2 \rangle$ as a function of centrality, and compared to
NA50 data~\cite{Abreu:2000xe,Topilskaya:2003iy} in Fig.~\ref{fig_pt2SPS}. 
Most of the observed $p_t$ dependence follows from the Cronin effect of 
the primordial component, represented by the dotted line.
The QGP suppression, which is stronger at high $p_t$ due to the increase
of the dissociation rate with $p$ (recall Fig.~\ref{fig_rate}), 
leads to a slight reduction of  $\langle p^2_t\rangle$, improving the 
agreement with data. The coalescence component is rather insignificant.  

\begin{figure}[!t]
\includegraphics[width=0.4\textwidth]{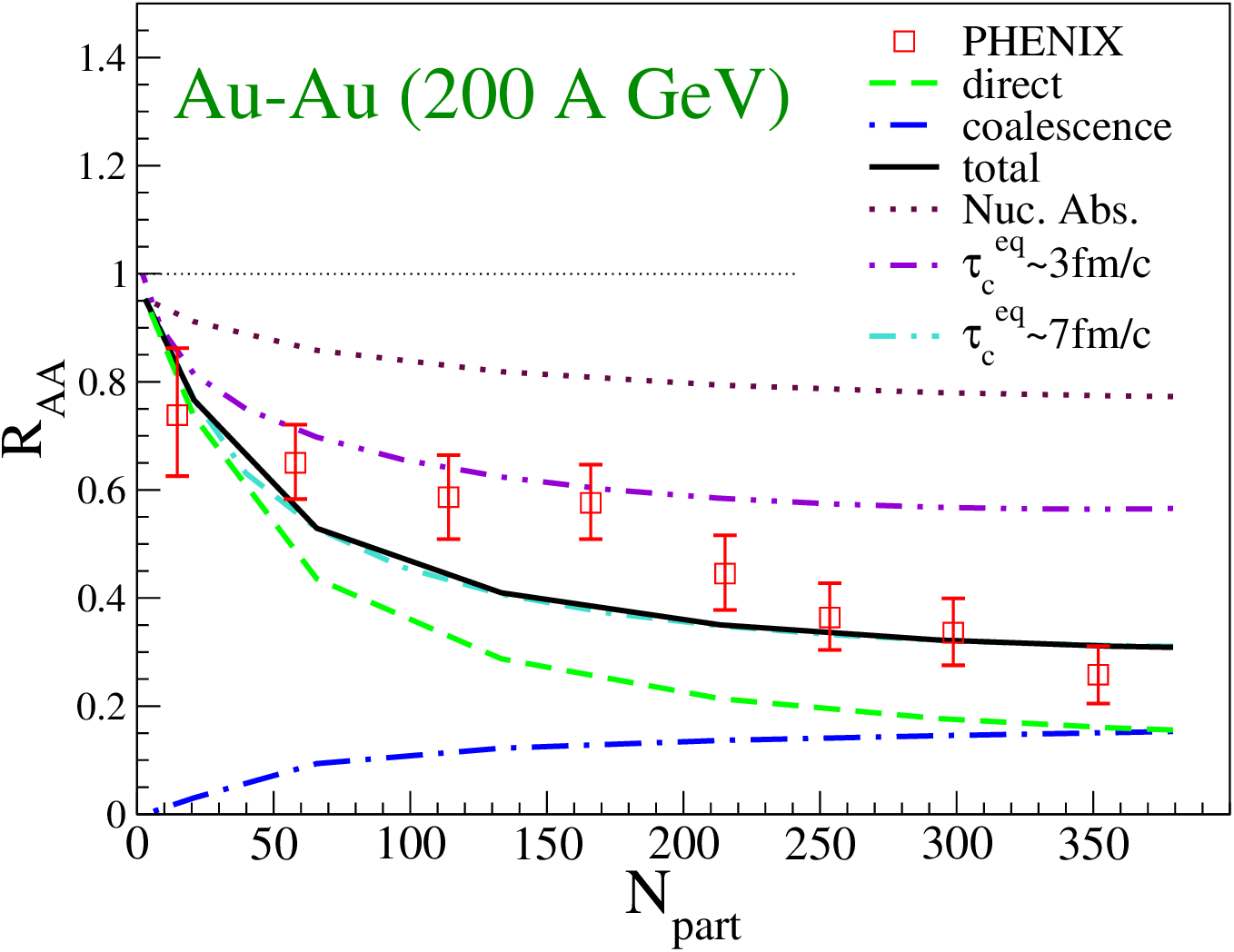}
\caption{(Color online). Results of the 2-component model for 
  $R_{AA}^{J/\psi}(N_{part})$ at RHIC, compared to PHENIX 
  data~\cite{Adare:2006ns}. The previous (momentum-independent) 
  model~\cite{Grandchamp:2003uw}, corresponds to the dash-double-dotted 
  line ($\tau^{\rm therm}_c$=3~fm/$c$) and the double-dash-dotted 
  ($\tau^{\rm therm}_c$=7~fm/$c$) line.
  The latter closely coincides with the total (solid line) using
  $p$-dependent rates (also with $\tau^{\rm therm}_c$=7~fm/$c$).  
  The dotted line represents the primordial input, while the dashed
  line additionally includes QGP and HG suppression. The coalescence 
  yield for $\tau^{\rm therm}_c$=7~fm/$c$ is given by the dash-dotted
  line.}   
\label{fig_RAARHIC}
\end{figure}
Next, we proceed to the centrality dependence of inclusive $J/\psi$
production in $Au$-$Au$ collisions at RHIC, as represented by the 
nuclear modification factor, $R_{AA}(N_{part})$ (the number of $J/\psi$'s
for a given number of participant nucleons, $N_{part}$, 
relative to that in $p$-$p$ collisions multiplied by the number of
binary collisions), cf.~Fig.~\ref{fig_RAARHIC}.
Previous $p$=0 results~\cite{Grandchamp:2003uw} with the updated nuclear
absorption cross section (but with identical coalescence contribution) 
overestimate the most recent PHENIX data for
$N_{part}$$>$200; increasing $\tau^{\rm therm}_c$ to 7~fm/$c$ improves 
this part at the expense of more peripheral collisions. 
The inclusion of the 3-momentum dependence (with $\alpha_s$=0.24)
does not resolve this potential discrepancy, despite the presence
of the leakage effect, for similar reasons as described above 
for SPS energies.  We note that the roughly equal partition of 
the 2 components for central collisions is quite similar to the results 
of Ref.~\cite{Yan:2006ve} (where the vacuum gluon dissociation mechanism
has been employed).  

\begin{figure}[!t]
\includegraphics[width=0.40\textwidth]{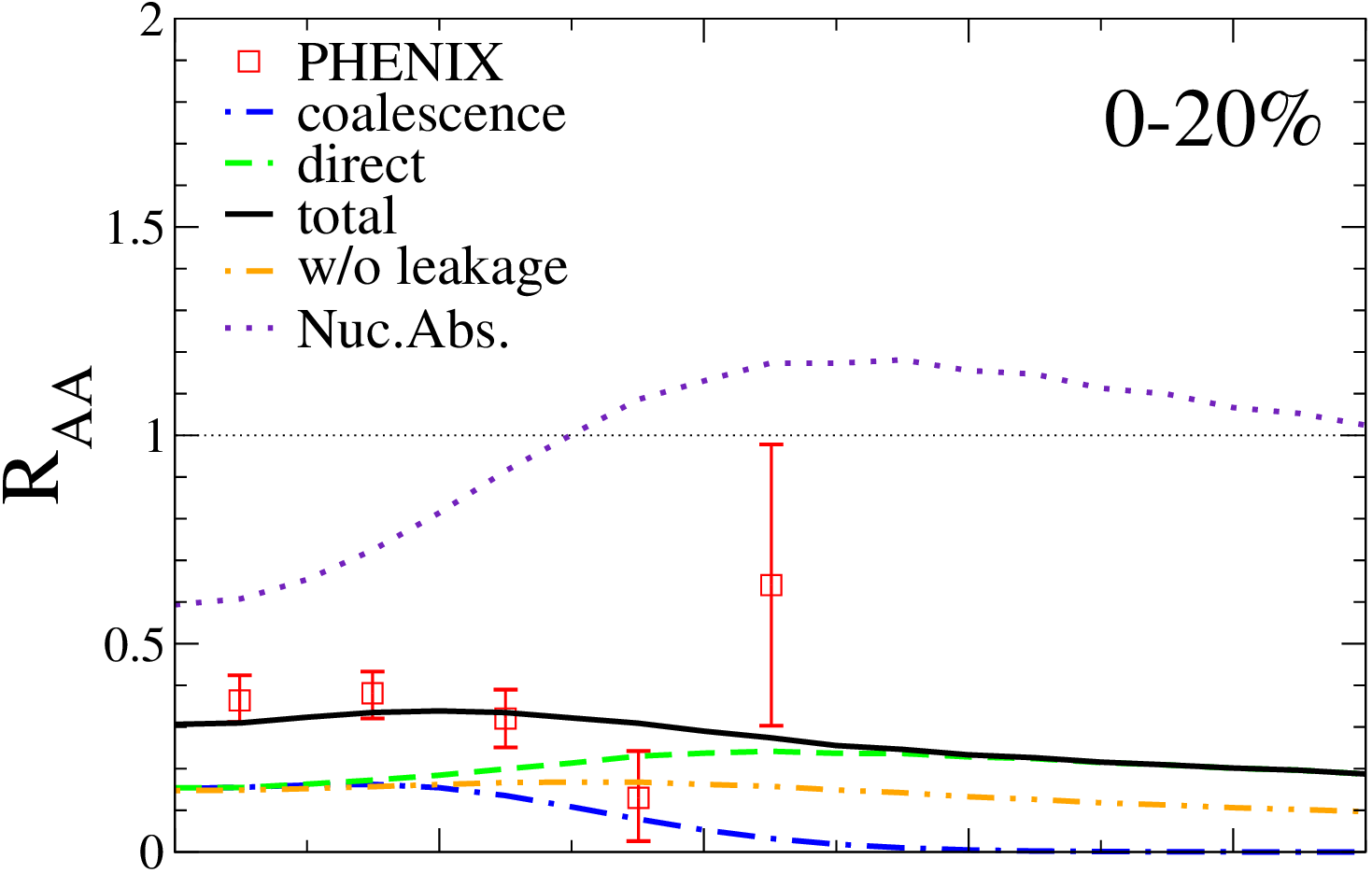}\\[1mm]
\includegraphics[width=0.40\textwidth]{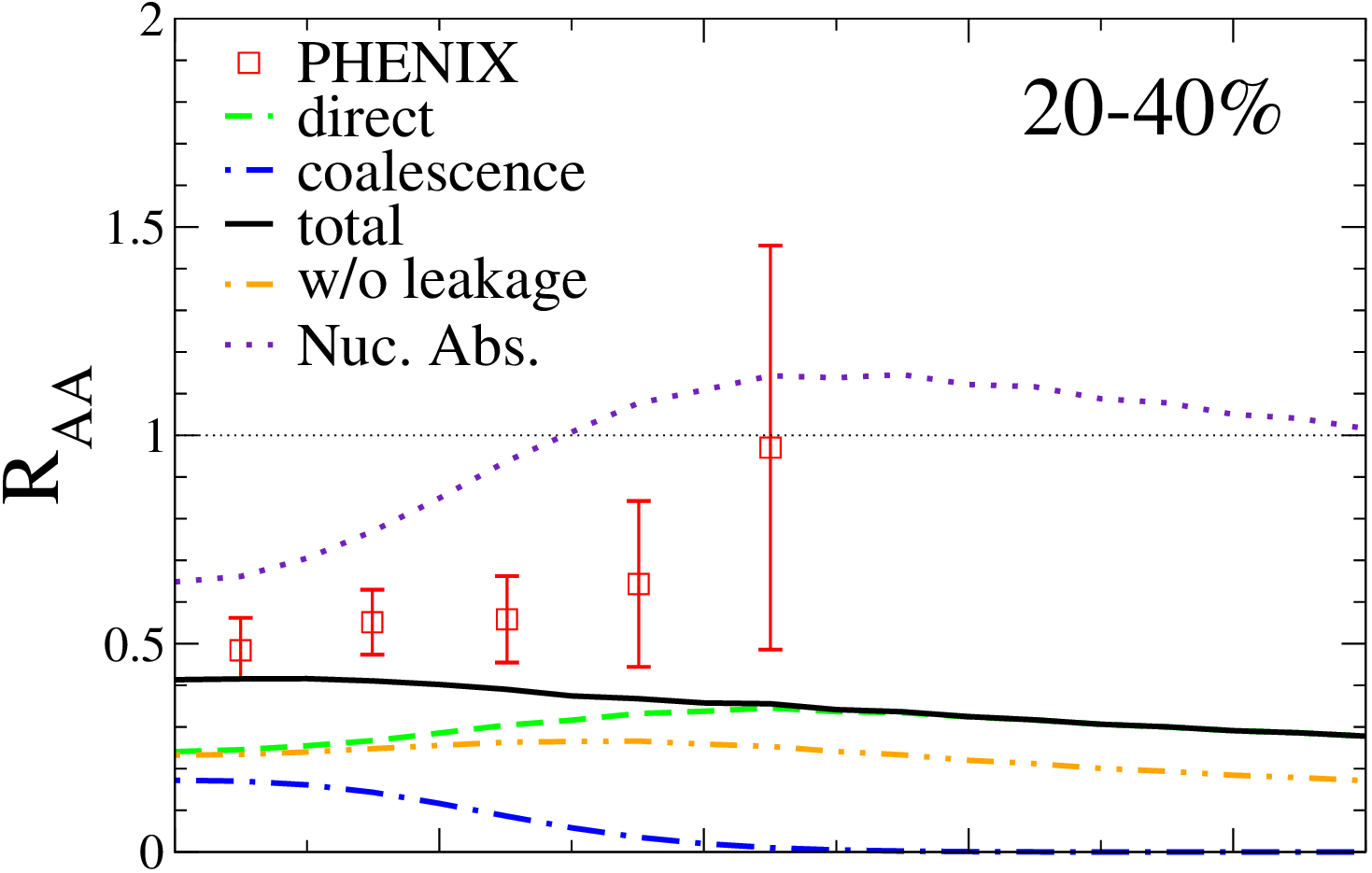}\\[1mm]
\includegraphics[width=0.40\textwidth]{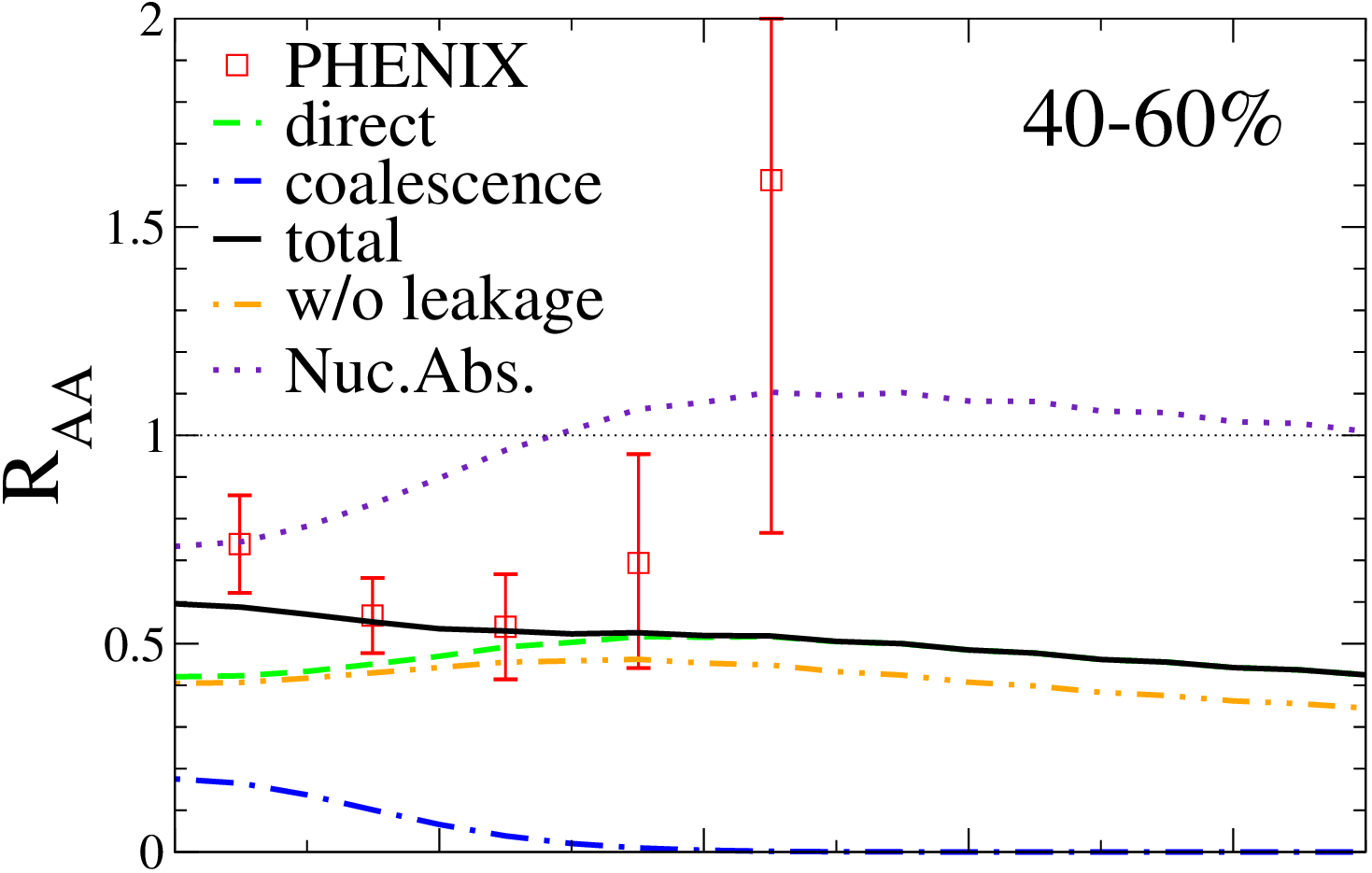}\\[1mm]
\includegraphics[width=0.40\textwidth]{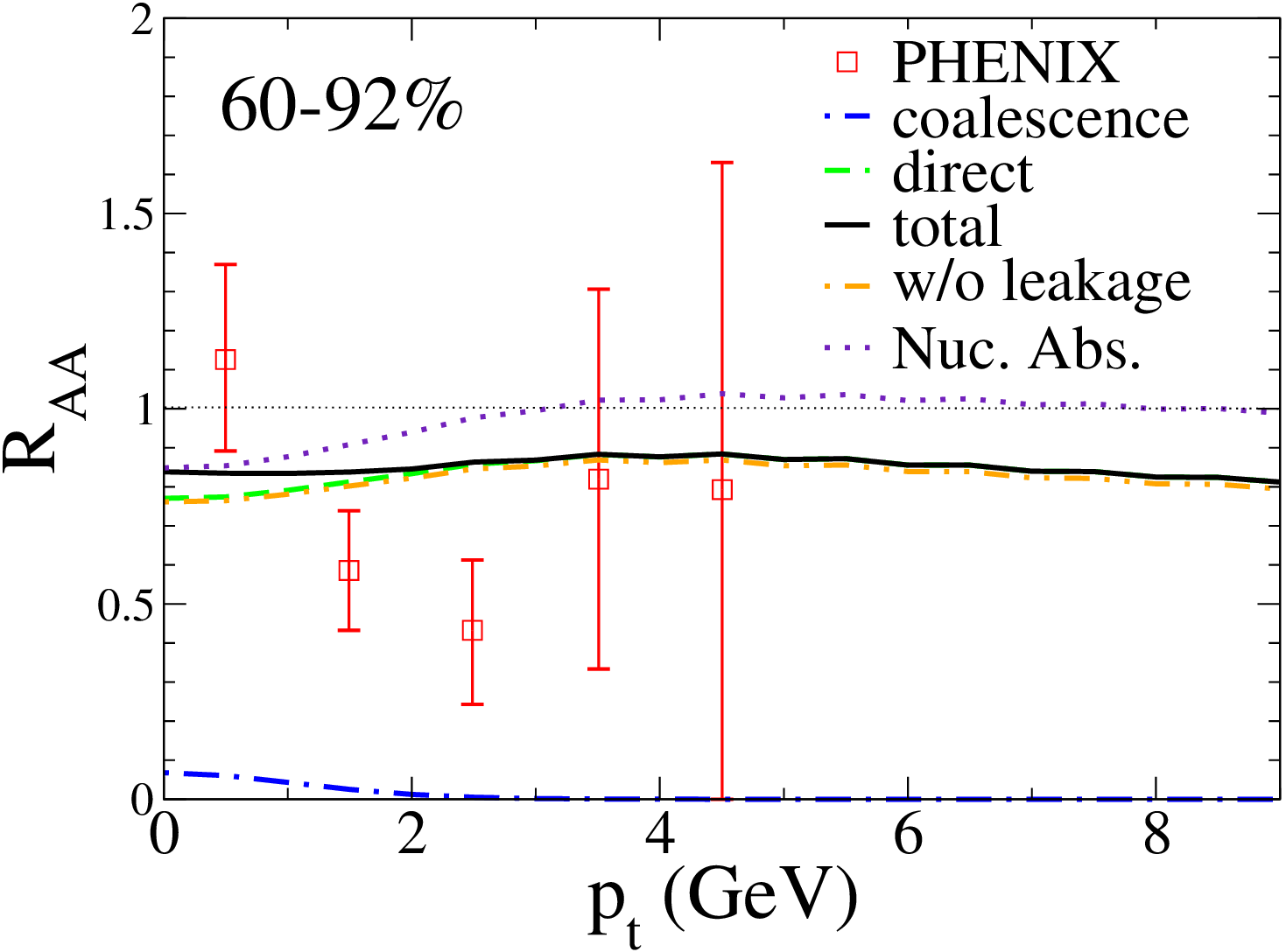}
\caption{(Color online). $R_{AA}$ vs. transverse momentum
  for different centrality selections of $Au$-$Au$ at RHIC. PHENIX 
   data~\cite{Adare:2006ns} are compared to our model calculations:
  initial primordial component (dotted line), including QGP and HG
  suppression (dashed line), and with leakage effect switched off
  (dash-double-dotted line); the coalescence contribution is 
  given by the dash-dotted line.}
\label{fig_RAAptRHIC}
\end{figure}
The key point is now how the inclusive centrality dependence of the 
two components reflects itself in the $p_t$ spectra. Our results for
$R_{AA}(p_t)$ for different centrality selections (approximated by
the average number of binary $N$-$N$ collisions) is compared to
PHENIX data~\cite{Adare:2006ns} in Fig.~\ref{fig_RAAptRHIC}.
As anticipated, the coalescence contribution is concentrated at
low $p_t$ (up to 2-4~GeV), most notably for central collisions but 
quickly ceasing for more peripheral ones. 
For the direct component, the $p_t$-broadening of the Cronin effect 
induces an appreciable rise of $R_{AA}(p_t)$ 
in the region from 2-5~GeV (cf.~the dotted lines).
This trend is largely counter-balanced by the QGP suppression  
(dash-double-dotted line, with leakage effect off, implemented
by ignoring the spatial fireball boundary), due to 
the increase of the quasifree dissociation rate with momentum.  
The leakage effect finally restores significant strength at higher
$p_t$ (up to $\sim$40\% for $p_t$$>$5~GeV) in the direct component 
(dashed lines).
The opposite $p_t$ dependence of the direct and coalescence spectrum 
combines into a rather flat total $R_{AA}(p_t)$ which is quite 
compatible with experiment. We emphasize that a proper description of
the absolute yields is an important ingredient to this finding (the
underestimate for 20-40\% central collisions could be improved
upon with a somewhat smaller $c$-quark thermalization time).   
{\it E.g.}, a pure coalescence spectrum can be compatible with
the central data, but it would be less convincing for more peripheral
collisions. Therefore, $R_{AA}(p_t)$ data may indeed discriminate
a two-component from a one-component model, especially if the
experimental uncertainty can be reduced.  

We have checked that within the current experimental accuracy of 
$R_{AA}(p_t)$ it is not possible to exclude different suppression 
mechanisms in the QGP medium. The data are also consistent with 
calculations employing a dissociation rate based on 
gluo-dissociation~\cite{Peskin:1979va} with vacuum binding energy and 
zero thermal gluon mass, since the relevant suppression regime
is for temperatures $T\le300$~MeV where the $J/\psi$ rate is only 
weakly momentum dependent. 

\begin{figure}[!t]
\includegraphics[width=0.4\textwidth]{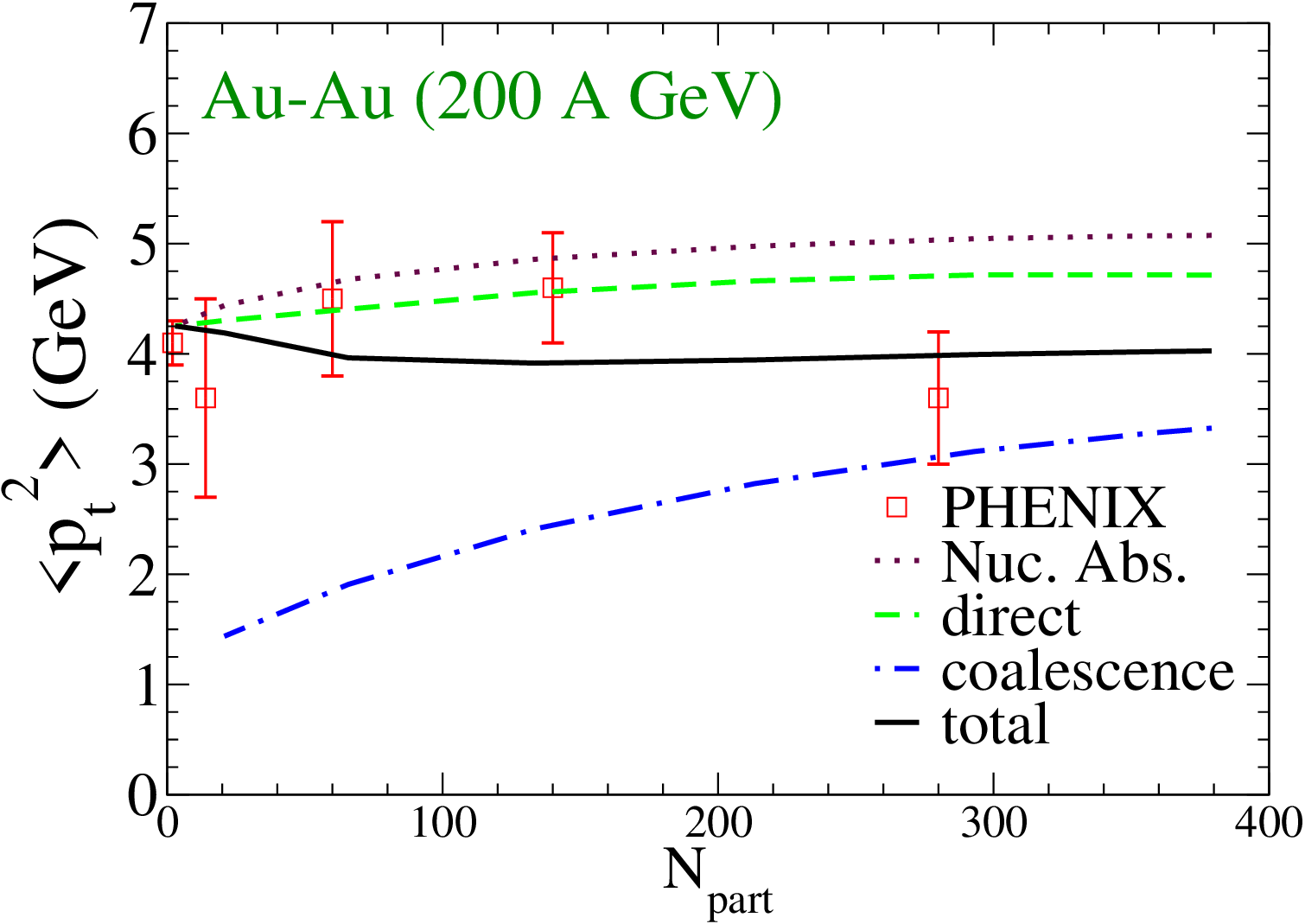}
\caption{(Color online). Centrality dependence of $\langle p^2_t\rangle$
  within the 2-component model at RHIC, compared to PHENIX 
  data~\cite{Adare:2006ns}. The dotted line corresponds to the 
  primordial input distribution (with nuclear absorption and Cronin 
  effect), the dashed line includes QGP and HG suppression, and 
  the dot-dashed line represents the coalescence component.
\label{fig_pt2RHIC}}
\end{figure}
We finally condense the $p_t$ spectra into a centrality dependence of
$\langle p^2_t\rangle$, as computed from our spectra and
compared to PHENIX data~\cite{Adare:2006ns} in Fig.~\ref{fig_pt2RHIC}.
This plot reiterates the importance of the soft coalescence spectra 
in central collisions to provide a near-flat centrality dependence. 
We recall, however, the currently
large uncertainty in the Cronin effect as inferred from $d$-$Au$ 
collisions. 

\section{Summary and Conclusions}
\label{sec_concl}
We have studied the 3-momentum dependence of $J/\psi$ production in
heavy-ion collisions based on a previously developed two-component
model which accounts for primordial production with subsequent
suppression and secondary regeneration close to the QCD 
phase boundary. For the direct component, we adopted a transport
approach including up-to-date empirical input for nuclear absorption 
and a Cronin effect in the initial state. The key microscopic ingredient 
is the charmonium dissociation rate in the QGP. We argued that the 
quasifree destruction mechanism provides a realistic treatment for 
small (in-medium) binding energies and the extension to finite 
3-momentum. For the coalescence component, we adopted a blast wave 
description at the hadronization transition within the same fireball
model used for the direct spectra. 
Our approach has essentially two parameters: the strong coupling constant
in the quasifree rate, which we adjust to the suppression in central 
$Pb$-$Pb$ at SPS, and the thermal relaxation time of $c$ quarks, which 
controls the magnitude of the coalescence component, adjusted to the
$J/\psi$ yield in central $Au$-$Au$ at RHIC. Within reasonable values
for these parameters, $\alpha_s$$\simeq$0.24 and 
$\tau_c^{\rm therm}$$\simeq$5-7~fm/$c$, an approximate overall 
description of the centrality dependence of inclusive $J/\psi$ 
production at SPS and RHIC emerges. 
The key point of our Letter is that, without further assumptions, the 
calculated $p_t$ spectra are largely consistent with available SPS and 
RHIC data. We argued that this supports the underlying 
momentum dependence of the dissociation rate in connection with 
reduced binding energies, as well as the presence of a $\sim$50\% 
coalescence contribution in central $Au$-$Au$ collisions at RHIC.  
More work is required to scrutinize these findings, \eg,
an extension to NA60 data at SPS, forward rapidities at RHIC,
predictions for LHC, as well as more accurate input from $d$/$p$-A
experiments. A microscopic transport treatment with $c$-quark spectra 
constrained by open-charm observables will be pursued and used to 
predict elliptic flow. Ultimately, the underlying charmonium properties
in the QGP should be consistent with lattice QCD results to establish
model-independent connections between the QCD phase diagram and
the matter created in heavy-ion collisions. 

\vspace{0.3cm}

\acknowledgments 
We are grateful to L.~Grandchamp for providing us with his codes, and to him and H.~van Hees for numerous 
helpful discussions. 
This work is supported by a US National Science Foundation CAREER 
award under grant No. PHY-0449489.

\end{document}